\documentclass[%
 aip,
 amsmath,amssymb,
 reprint,%
]{revtex4-1}

\usepackage{graphicx}
\usepackage{dcolumn}
\usepackage{bm}

\usepackage[utf8]{inputenc}
\usepackage[T1]{fontenc}
\usepackage{mathptmx}

\usepackage[english]{babel}
\usepackage[utf8]{inputenc}
\usepackage[colorinlistoftodos, color=green!40, prependcaption]{todonotes}
\usepackage[pdftex, pdftitle={Article}, pdfauthor={Author}]{hyperref} 

\begin{document}


\title{Success rate analysis of the response of an excitable laser to periodic perturbations}

\author{Jordi Tiana-Alsina}
    \email[Correspondence email address: ]{jordi.tiana@upc.edu}
    \affiliation{Department of Physics, Universitat Polit\`ecnica de Catalunya, Rambla St. Nebridi 22, Terrassa 08222, Barcelona, Spain}
    
\author{Bruno Garbin}
    \affiliation{Université Paris-Saclay, CNRS, Centre de Nanosciences et de Nanotechnologies, 91120, Palaiseau, France}
    
\author{Stephane Barland}
    \affiliation{Université Côte d’Azur, CNRS UMR 7010, Institut de physique de Nice, 1361 Route des Lucioles, F-06560 Valbonne, France}

\author{Cristina Masoller}
    \email[Correspondence email address: ]{cristina.masoller@upc.edu}
    \affiliation{Department of Physics, Universitat Polit\`ecnica de Catalunya, Rambla St. Nebridi 22, Terrassa 08222, Barcelona, Spain}


\begin{abstract}
We use statistical tools to characterize the response of an excitable system to periodic perturbations. The system is an optically injected semiconductor laser under pulsed perturbations of the phase of the injected field. We characterize the laser response by counting the number of pulses emitted by the laser, within a time interval, $\Delta T$, that starts when a perturbation is applied. The success rate, $SR(\Delta T)$, is then defined as the number of pulses emitted in the interval $\Delta T$, relative to the number of perturbations. The analysis of the variation of $SR$ with $\Delta T$ allows to separate a constant lag of technical origin and a frequency-dependent lag of physical and dynamical origin. Once the lag is accounted for, the success rate clearly captures locked and unlocked regimes and the transitions between them. {We anticipate that the success rate will be a practical tool for analyzing the output of periodically forced systems, particularly when very regular oscillations need to be generated via small periodic perturbations}.  
\end{abstract}


\maketitle
\begin{quotation}
Excitable systems, where small perturbations produce almost no response but large enough perturbations do, are ubiquitous in nature. Examples include neurons and cardiac cells. It is important to understand how excitable systems respond to periodic perturbations, and to characterize their locked and unlocked dynamical behaviours. Here we study experimentally an optically injected laser that has been shown to be excitable (a large enough perturbation triggers the emission of a pulse of light). We analyze how the laser responds to a periodic perturbation of the injected field using a statistical analysis tool (referred to as success rate) which uncovers a lag in the laser response that is traced back to the experimental conditions. Once this lag is corrected, the success rate analysis unveils different locked regimes; in these regimes, $n$ light pulses are emitted every $m$ perturbations.
\end{quotation}

\section{Introduction}
Controlling a stochastic excitable system with periodic perturbations is a challenging problem with applications across disciplines. A practical important example is that of an artificial pacemaker, which delivers electrical impulses to regulate the function of the heart. Interference effects, due to different oscillation or response times, lead to dynamical regimes in which a system can oscillate with frequency $\omega_s$ when it is periodically perturbed with frequency $\omega_f$. The system is said to be $m:n$ locked, when it performs $n$ oscillations every $m$ perturbations. As the perturbation frequency or the natural rhythm of the system vary, transitions between different locked and unlocked states occur~\cite{Pikovsky_2003}. In the unlocked states the system' dynamics is either oscillatory with incommensurate frequencies, or chaotic~\cite{glass_1982,piro_1998}. 

Locking is often difficult to identify and quantify, particularly in the presence of noise. Methods based on the analysis of the time intervals between consecutive responses, spectral or correlation analysis can identify the locking regions, but are unsuitable for quantifying the strength of the locking (the regularity of the system's response), and therefore, they are unable to provide a systematic way to identify the optimal locking conditions. Recently, some of us have proposed a methodology based on the so-called ``success rate'' (SR)~\cite{Tiana-Alsina_2018, Tiana_pre_2019}. Using a semiconductor laser with external optical feedback, whose pump current was periodically modulated, we have shown that SR analysis allows to identify the modulation waveform that provides the most robust locked conditions~\cite{Tiana-Alsina_2018}, and to quantify the strength of the locking of the laser intensity to the current modulation~\cite{Tiana_pre_2019}. 

{{To demonstrate the general applicability of this methodology, in~\cite{Tiana_pre_2019} we also analyzed simulations of an stochastic bistable system under square-wave forcing. We showed that spectral analysis, correlation analysis, and the analysis of the distribution of residence times in each state allow to identify the locking regions, but do not provide a precise way to quantify the degree of locking (see the supplementary information of~\cite{Tiana_pre_2019}). While the theory underlying the SR approach needs to be elaborated, we speculate that the success of this technique is due to the fact that both, the input signal and the output signal are transformed into point processes, and only the times when the perturbations occur and the times when the responses occur are taken into account. Therefore, the method is intrinsically nonlinear.}}

Here we apply this methodology to a well-known laser system that can display locked behaviour: an optically injected semiconductor laser. Under constant injection conditions the laser displays different dynamical regimes, including the so-called injection-locking (where the laser emits a constant output whose wavelength is identical to that of the injected light), periodic oscillations, and chaotic oscillations~\cite{review,ohtsubo_book,marc,yanhua,huyet1,huyet2,kovanis,ana}, with or without extreme fluctuations~\cite{prl_2011,zamora_2013,marita}. When the phase of the injected field is perturbed, under appropriate conditions the laser emits pulses that are locked to the phase perturbations, and whose excitable nature was demonstrated in~\cite{Barland_PRE_2013,Bruno_2017}. In~\cite{Barland_PRE_2013} it was shown that there is a perturbation threshold beyond which the response of the laser is independent of the strength of the perturbation. When multiple perturbations are applied in short time frame, this system can display a refractory period~\cite{Bruno_2017}, during which perturbations are not able to elicit a response or a more complex resonator behavior and multipulse excitability \cite{multipulse,dolcemascolo2018resonator}. This architecture also supports interesting neuromorphic applications such as memories or inter-neuron communications \cite{garbin2015topological,robertson2019toward}. When additional physical mechanisms are taken into account beyond only field and carrier dynamics, other types of excitable dynamics may even arise, as in \cite{dillane2019square,dillane2019neuromorphic}. In presence of periodic parameter modulation, a devil's staircase was recently observed \cite{lingnau2019devil}. One of the key issues for understanding those complex dynamical regimes is to use appropriated tools for their characterization. Here we use SR analysis to characterize, in the excitable regime, the response of the laser to periodic phase perturbations. We identify a lag in the response that varies with the frequency of the phase perturbations, and once this lag is taken into account, SR analysis identifies the perturbation frequencies that produce $m:n$ locked laser pulses.

This paper is organized as follows. Section~\ref{sec:exp} describes the experimental setup; Sec.~\ref{sec:methods} describes the SR methodology, Sec.~\ref{sec:results} presents the results; Sec.~\ref{sec:dis} presents the discussion and Sec.~\ref{sec:conc} summarizes the conclusions. 

\section{Experimental setup}\label{sec:exp}

\begin{figure}[tb]
\begin{center}
\includegraphics[width=1\columnwidth]{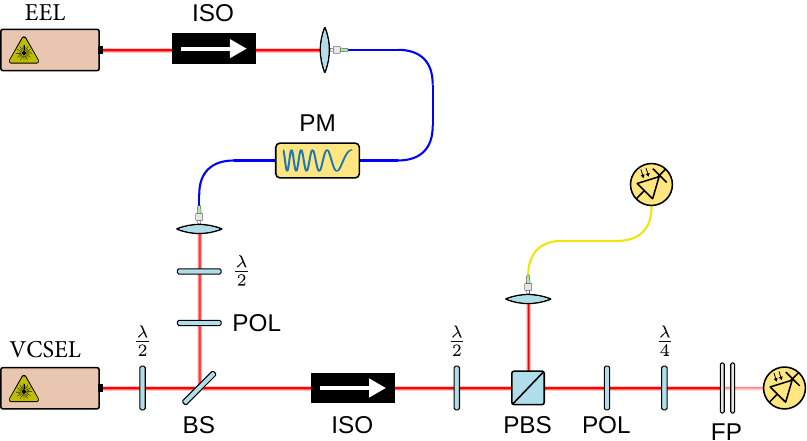}
\caption{Schematic of experimental setup. A phase modulator (PM) perturbs the phase of the field emitted by a tunable edge emitting laser (EEL) that is injected into a VCSEL laser. ISO, optical isolator; $\lambda/2$, half-wave plate; $\lambda/4$, quarter-wave plate; POL, polarizer; PBS, polarizing beam splitter; FP, Fabry-P\'erot interferometer. }
\label{fig:Fig_ES}
\end{center}
\end{figure}

The experimental setup is depicted in Fig.~\ref{fig:Fig_ES}. It consists of a single transverse mode Vertical Cavity Surface Emitting Laser (VCSEL) locked to a coherent external field (driving beam).
The VCSEL is operated well above the threshold (the pump current is 1.3~mA and the threshold is about 0.2~mA). The driving beam is produced by a tunable edge emitting laser (EEL). In order to ensure unidirectional coupling, a 40~dB isolator is placed right after the output of the EEL. The beam is then focused into a single mode fiber which guides it to a fiber-coupled 10~GHz Lithium Niobathe phase modulator. The beam is then collimated at the output of the modulator and steered by mirrors towards the VCSEL. Before the final steering beam splitter, a half-wave plate and a vertical polarizer allow to control the injection strength. The VCSEL output is collimated and its (single-) polarization state is rotated to match that of the driving beam via a half-wave plate. The beam transmitted through the final steering beam splitter is sent to a 30dB optical isolator and split via a polarizing beam splitter. One part is then sent to a Fabry Perot interferometer while the other is injected in a 9.5~GHz fiber-coupled optical detector. The ratio between both detection paths can be adjusted thanks to a half-wave plate. 

We bring the VCSEL into an excitable state by tuning the optical frequency difference between the injected field and the VCSEL field. This is done by adjusting the VCSEL pump current.

In order to apply periodic phase perturbations, the phase modulator is driven by a pulse generator whose output is triggered by a sinusoidal signal. In this way, the phase of the injected field is periodically perturbed by the pulse generator that applies 100~ps jumps of $\sim\pi$ amplitude, which are triggered by a sinusoidal signal whose frequency is varied in the range 0.5~GHz-6.5~GHz. 

\section{Methods}\label{sec:methods}
\begin{figure}[tbp]
\begin{center}
\includegraphics[width=1\columnwidth]{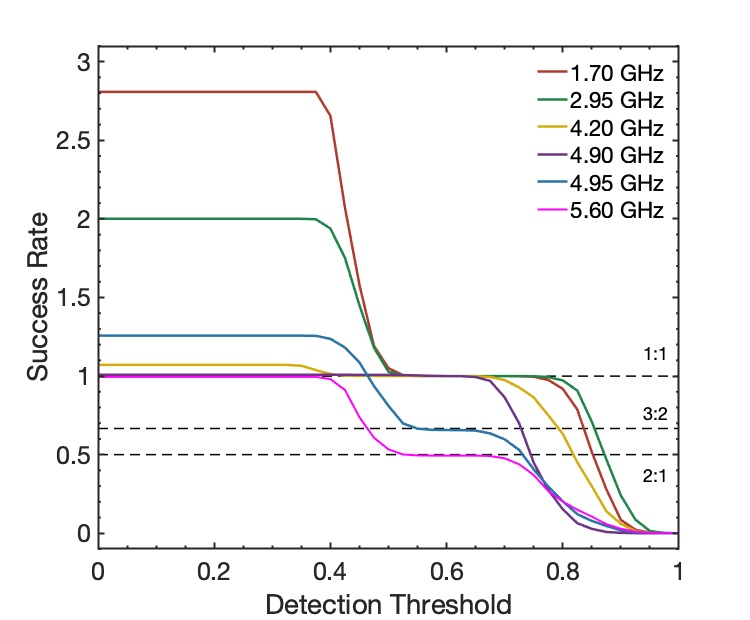}
\caption{Analysis of the role of the threshold used to detect the laser pulses. The success rate is plotted vs. the threshold for different perturbation frequencies that produce locking $1$:$1$, $3$:$2$ and $2$:$1$. {{The detection threshold is normalized between 0 and 1, corresponding to the lowest and highest intensity value respectively.}}}
\label{fig:Fig_Th}
\end{center}
\end{figure}

The success rate ($SR$) measures the response of the laser per modulation cycle: if the laser emits one pulse per cycle, $SR=1$, if it emits one pulse every two cycles, $SR=0.5$, etc. A drawback of this definition is that it does not take into account the regularity of the timing of the pulses: the pulses could be emitted at any phase of the modulation cycle, or at a well-defined phase. Therefore, we consider a detection window, $\Delta T$, that starts at each maximum of the sinusoidal, and count only the pulses that are emitted within this time interval. Then, the success rate is a function of $\Delta T$~\cite{Tiana-Alsina_2018}:
\begin{equation}
\mathrm{SR}(\Delta T) =\frac{\mbox {\# of pulses }}{\mbox {\# of perturbations} }.
\label{eq:SR}
\end{equation}
To detect the laser pulses we have used the Matlab function "findpeaks" to analyze the intensity time series. This function finds the local maximums (peaks) of a signal that are above a given threshold. Unless otherwise specified, the threshold used is $0.6$ (the role of this threshold will be discussed in Fig.~\ref{fig:Fig_Th}). To filter out noisy fluctuations only peaks whose prominence is larger than $0.3$ were counted. 

\begin{figure}[tbp]
\begin{center}
\includegraphics[width=1.08\columnwidth]{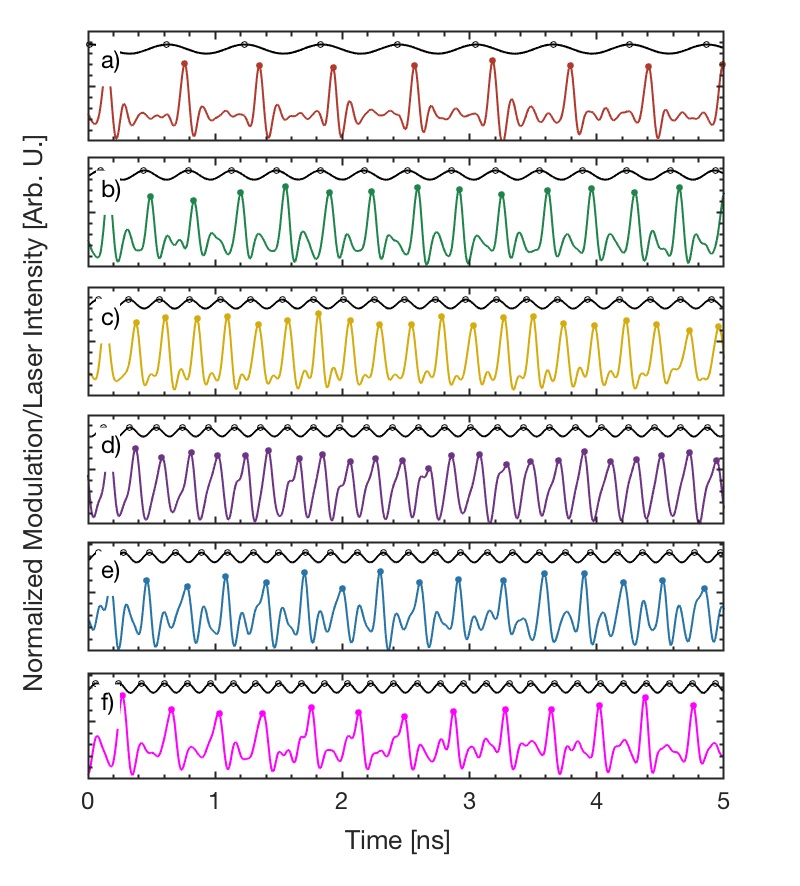}
\caption{Laser intensity time series (color lines) and sinusoidal signal that modulates the pulse generator that, in turn, produces $\pi$ perturbations in the phase modulator (black lines). In (a), (b), (c) and (d) the laser emits one pulse after each perturbation (locking $1$:$1$; we note that in (a)-(c) pulses are followed by smaller relaxation oscillations); in  (e) there are two pulses every three perturbations (locking $3$:$2$) and in (f), there is one pulse every two perturbations (locking $2$:$1$). The modulation frequency is $1.70$~GHz (a), $2.95$~GHz (b), $4.20$~GHz (c), $4.90$~GHz (d), $4.95$~GHz (e) and $5.60$~GHz (f). In all the panels the intensity time series has been normalized between 0 and 1 and lagged by $5$~ns to account for the delay in signal transmission and laser response (see text for details).}
\label{fig:Fig_TS}
\end{center}
\end{figure}

\begin{figure}[tbp]
\begin{center}
\includegraphics[width=1.0\columnwidth]{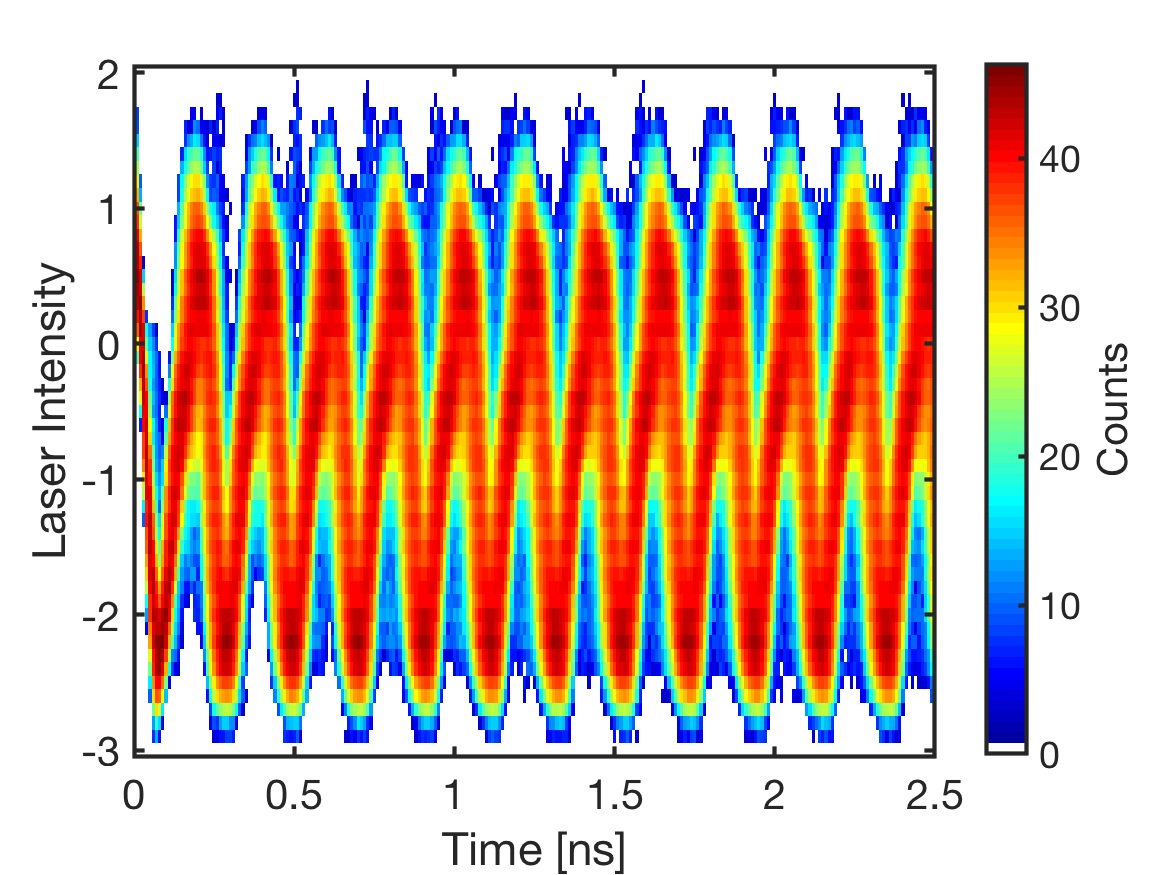}
\includegraphics[width=1.0\columnwidth]{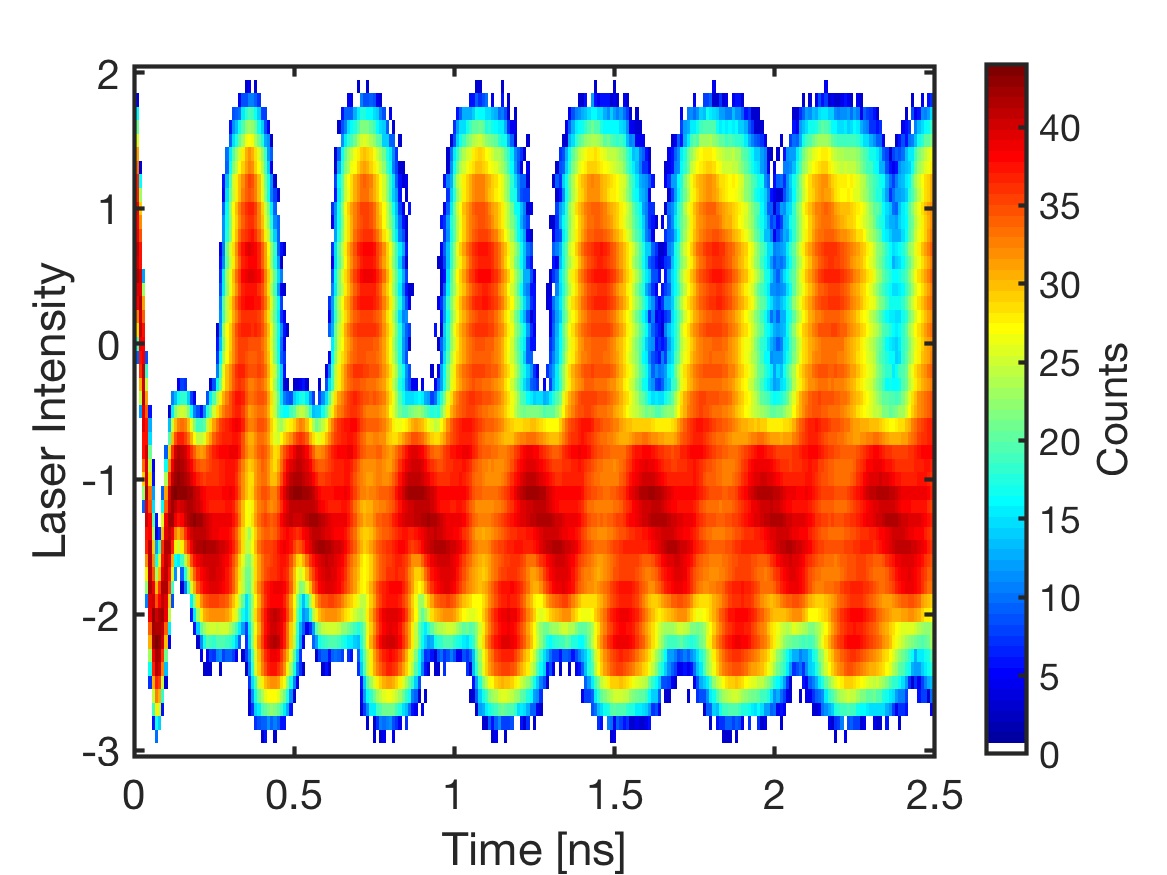}
\caption{Superposition of $4 \times 10^5$ pulses when the sinusoidal signal that modulates the pulse generator has a frequency of 4.90~GHz (a), 5.60~GHz (b).}
\label{fig:time_traces}
\end{center}
\end{figure}

\begin{figure}[tbp]
\begin{center}
\includegraphics[width=1.0\columnwidth]{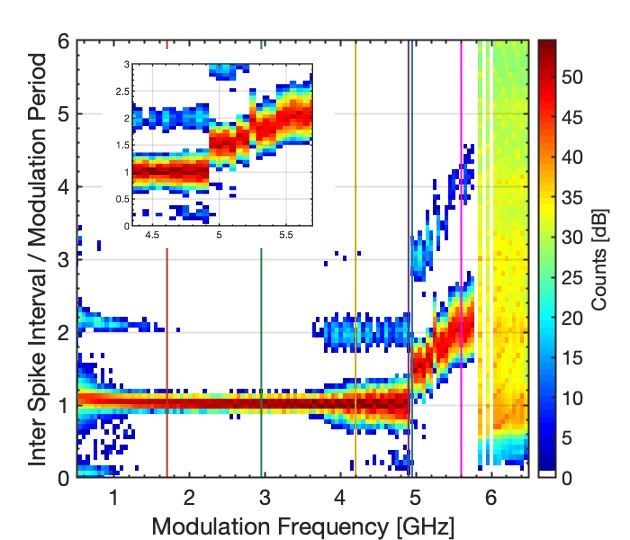}
\caption{Distribution of time intervals between pulses (inter-spike interval ISI distribution) in color code vs. the modulation frequency. In the vertical axis the intervals are normalized by the modulation period. In order to  enhance the plot contrast, the color scale indicates the logarithmic of the number of intervals (the white color stands for zero counts). The vertical lines indicate the frequencies used in Fig.~\ref{fig:Fig_TS}; the inset shows in detail the transitions from $1$:$1$ to $3:2$ and  $2$:$1$ locking.}
\label{fig:Fig_ISI}
\end{center}
\end{figure}

\begin{figure}[tbp]
\begin{center}
\includegraphics[width=0.85\columnwidth]{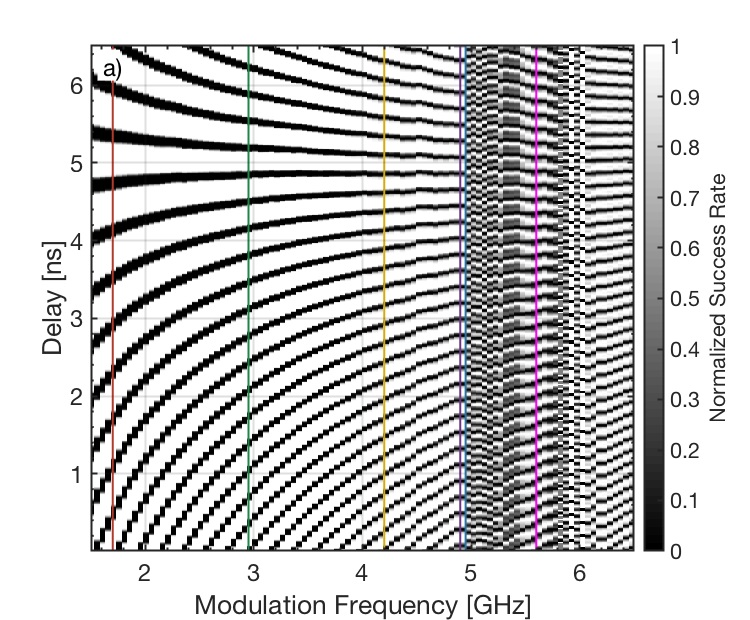}
\includegraphics[width=0.90\columnwidth]{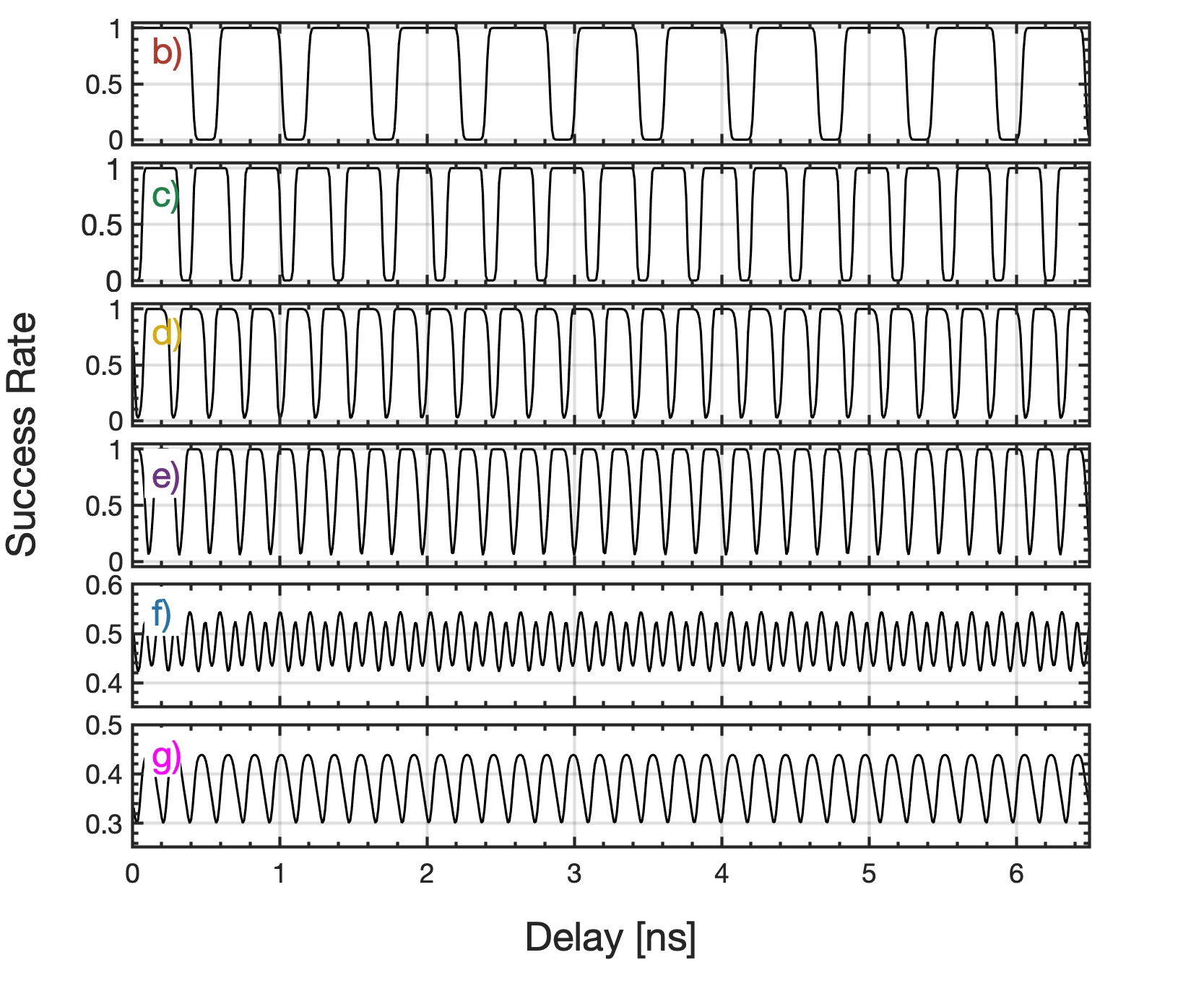}
\caption{{{Demonstration of a delay between electronics and optics.}} (a) Normalized success rate as a function of the perturbation frequency and the delay between the input and output signals (the sinusoidal that triggers the pulse generator, and the laser intensity). For a given frequency, the success rate (SR) oscillates with the delay between two values that have been normalized to 0 and 1. {{We note that when the lag is about 5~ns (dashed horizontal line), the SR is maximum for all frequencies.}} Panels (b)-(g) display the actual SR values. In (a) the vertical lines indicate the frequencies used in (b)-(g), which are as in Fig.~\ref{fig:Fig_TS}: $1.70$~GHz (b), $2.95$~GHz (c), $4.20$~GHz (d), $4.90$~GHz (e), $4.95$~GHz (f) and $5.60$~GHz (g). The SR is calculated using a detection window $\Delta T$ = $70\%$ of the perturbation period.}
\label{fig:Fig_delayMap}
\end{center}
\end{figure}

\begin{figure}[tbp]
\begin{center}
\includegraphics[width=0.95\columnwidth]{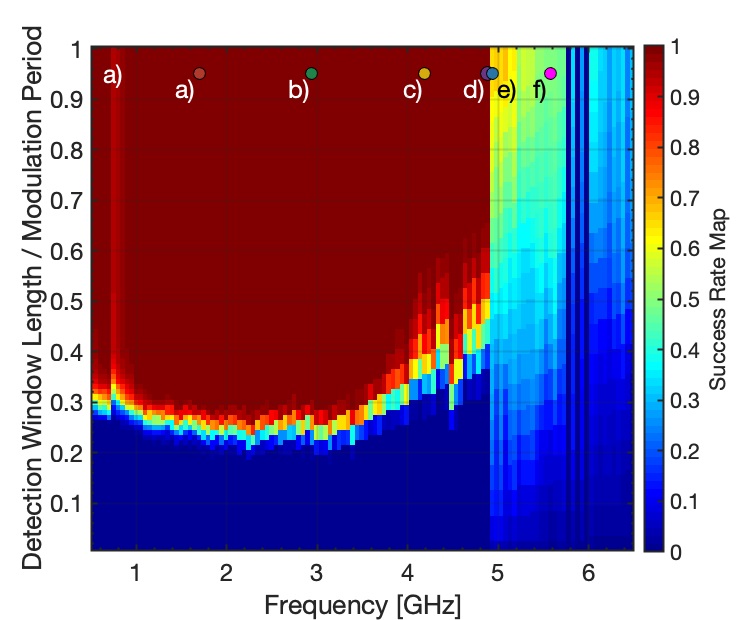}
\includegraphics[width=0.95\columnwidth]{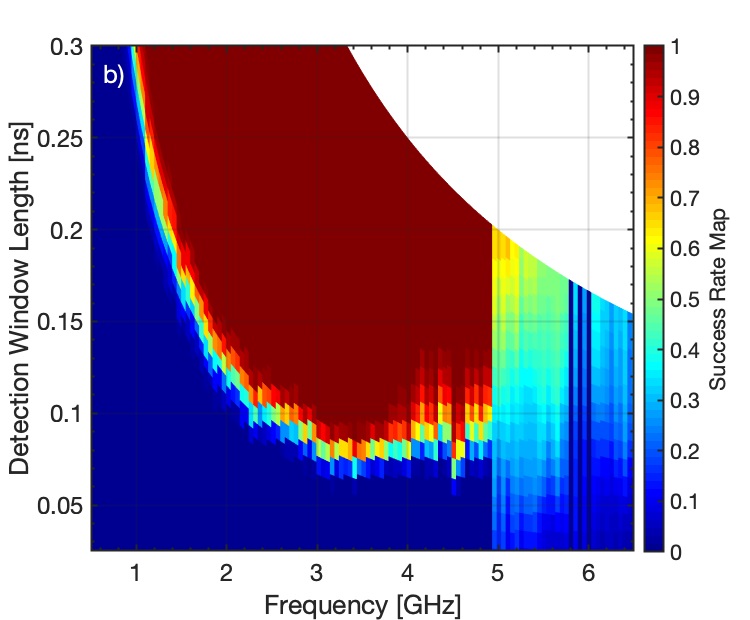}
\caption{Success rate (in color code) as a function of the perturbation frequency and the duration of the detection window. To calculate the SR the response signal (laser intensity) has been delayed 5 ns with respect to the input signal (sinusoidal signal). In (a) the duration of the detection window is normalized to the period of the sinusoidal signal (i.e., the period of the phase perturbations) while in (b) it is shown without normalization. The vertical lines in (a) indicate the frequencies used in Fig.~\ref{fig:Fig_TS}.}
\label{fig:Fig_SRmap}
\end{center}
\end{figure}

\section{Results}\label{sec:results}

Figure~\ref{fig:Fig_TS} displays, for different modulation frequencies, the dynamics during 5 ns (the laser intensity is shown in color line, and the sinusoidal signal that triggers the pulse generator that applies a perturbation to the phase modulator, in black line). As will be explained below, there is a 5~ns lag between the input signal (sinusoidal modulation) and the output signal (laser intensity), which has been corrected: here the intensity time series has been lagged 5~ns with respect to the sinusoidal. In order to illustrate the long term behaviour, Fig.~\ref{fig:time_traces} displays the superposition of a large number of intensity pulses ($4 \times 10^5$ modulations periods are shown, the origin of time is set when the laser emits an spike). 

In Fig.~\ref{fig:Fig_TS}(a-d) locking $1$:$1$ (number of cycles of the input signal : number of emitted pulses or spikes) is observed. For low modulation frequencies relaxation oscillations occur after each induced spike (panels (a-c)); for higher frequencies the laser responds to each perturbation with just one spike (panel (d)). However, if the perturbations become too fast the laser can not follow them and a transition to a different locking regime occurs. Specifically, when the frequency is higher than $f_{mod}=4.95$~GHz, $3$:$2$ locking occurs (panel (e)), and for even higher frequency, there is locking $2$:$1$ (panel (f)).

The different locking regimes can be characterized in terms of the distribution of times between consecutive pulses (the so-called inter-spike-interval ISI distribution). Figure~\ref{fig:Fig_ISI} displays the ISI distribution (in color code) vs. the frequency of the perturbations. The intervals in the vertical axis are normalized to the perturbation period, $T_{mod}= 1/f_{mod}$, and the histograms are computed with bins centered at $n_{Tmod}$. For frequencies in the range $f_{mod}=1-4$~GHz the ISI histograms show a single flat narrow line centered at one, which indicates that the laser emits one and only one pulse per perturbation. For frequencies below 1~GHz or in the range 4-4.93~GHz we observe weak side peaks (note the logarithmic color scale), which indicate that the pulses do not always follow the perturbation (some ISIs are either smaller or longer than the perturbation period).

At $f_{mod}=4.95$~GHz there is an abrupt transition after which, through a sequence of step-like plateaus (see inset) a regime where the laser emits a spike every two perturbations is reached ($f_{mod}=5.25$~GHz). This denotes the transition from locking 1:1 to locking 2:1. Within this transition noisy locking 3:2 (where two pulses are emitted every three perturbations) occurs (see in the inset the plateau at $1.5=3/2$). In the inset, for higher frequencies, a small plateau at $ISI/T_{mod}=2$ is observed for modulation frequencies above 5.5~GHz. For even higher frequencies the laser is unlocked and the ISI distribution is broad. {{The abrupt transition and the plateaus are also seen when different thresholds are used to detect the laser pulses, the main difference being that for higher thresholds some pulses are not detected and the ISI distribution has longer intervals, while for lower thresholds, smaller oscillations (relaxation oscillations) are detected and the ISI distribution has shorter intervals.}}

To analyze the statistical properties of the times when the laser pulses are emitted in relation to the times when the phase perturbations are applied, we calculate the success rate (SR) as defined in Sec. \ref{sec:methods}. %

The distribution of interspike time intervals (shown in Fig.~\ref{fig:Fig_ISI}), indeed, does not convey information about the timing of the spikes with respect to the timing of the perturbations. A broad peak in the ISI distribution, as observed in Fig.~\ref{fig:Fig_ISI} for frequencies between 4 and 5~GHz, indicates that there is a dispersion in the relative timing. Moreover, the time lag between the input signal (sinusoidal modulation) and output signal (laser intensity) can not be inferred from the ISI distribution. In fact, in our experimental setup there is a very long delay (of several nanoseconds) of purely technical origin between the time in which the sinusoidal signal reaches a maximum and the time at which the laser is perturbed. This delay, $\tau_{exp}$, is due to the propagation of the electronic signals in radio-frequency cables, the response time of both the RF amplifier and the phase modulator, and finally optical propagation of the driving beam from the phase modulator to the VCSEL. The presence of this delay can not be inferred from Fig.~\ref{fig:Fig_ISI}. One could try to recover the value of $\tau_{exp}$ from dedicated measurements done in a parameter region in which the VCSEL responds linearly to phase perturbations (assuming that such linear regime exists); however, an alternative practical approach is to recover the value of $\tau_{exp}$ using the success rate (SR) quantifier.

To that aim, we calculate the SR when delaying the response time trace (the laser intensity) with respect to the input signal (sinusoidal modulation). The results are presented in Fig.~\ref{fig:Fig_delayMap}. 
For a given frequency, as the delay varies the SR oscillates between two values (similar results are obtained when plotting the cross-correlation, however, the variation of the cross-correlation, not shown, is smoother). In the top panel the two values have been normalized to 0 and 1 for better visualization. We observe that when the lag is {{about}} 5~ns, the SR is maximum for all frequencies. 

Figure~\ref{fig:Fig_SRmap} displays the SR in color code as a function of the detection time window, $\Delta T$, and the modulation frequency [in panel (a) $\Delta T$ is normalized to the modulation period, while in panel (b) is not]. In both plots the SR was calculated after correcting for the delay identified in the previous analysis, i.e., after lagging the response signal (laser intensity) by 5~ns with respect to the input signal (sinusoidal modulation that triggers the pulse generator). 

In addition to allowing us to identify the locking regions, these plots allow us to assess the response time of the laser under different perturbation frequencies. For low frequencies (from $f_{mod}=0.5$ to $3.5$~GHz) the SR is equal to zero if $\Delta T/T_{mod} < 0.3$, and is equal to one otherwise. This ``flat'' (0 or 1) behavior reveals that there is a well-defined response time: the laser emits a pulse $0.3 T_{mod}$ after each maximum of the sinusoidal signal (we remark that in Fig.~\ref{fig:Fig_SRmap} the signal transmission lag $\tau_{exp}=5$ ns has been corrected). In other words, in this frequency range the response time increases with the period of the perturbation.
For perturbation frequencies in the range of $f_{mod}=3.5-5$~GHz the laser response is about $0.1$~ns, and is rather independent of $T_{mod}$ (see Fig. ~\ref{fig:Fig_SRmap}(b)).
For modulation frequencies above $f_{mod}=5$~GHz there is not a well defined response time.

Finally, we address the role of the threshold used to detect the spikes.
Figure~\ref{fig:Fig_Th} displays the SR as a function of the detection threshold for the same perturbation frequencies as in Fig.~\ref{fig:Fig_TS}. Two plateaus are observed for thresholds around $0.2$ and $0.6$. The first plateau shows values of SR larger than one because the low spike detection threshold used counts the relaxation oscillations that follow each pulse as pulses induced by the phase perturbations.

\section{Discussion}\label{sec:dis}

In contrast to cross-correlation analysis, SR analysis is a nonlinear method that is appropriate for characterizing systems whose output signals can be reduced to a set of event occurrence times that form a point process. Here, the details of the input signal (sinusoidal modulation that triggers phase perturbations) and the output signal (the laser intensity) have been disregarded as both signals have been transformed into point processes, and only the times when the perturbations are triggered, and the times when the laser pulses are emitted, have been analyzed. 

In Fig.~\ref{fig:Fig_SRmap} we see that the variation of the SR with the perturbation frequency is nontrivial. It can be interpreted as follows. When computing the SR, one actually measures the delay between one perturbation and the \textit{first following response}. If the time response of the system is longer than the perturbation period, the measured response may in fact have been caused by an earlier perturbation. Therefore, one actually measures the \textit{remainder} of the total response time divided by the perturbation period. Oscillations in this remainder reveal that the total response time is larger than the modulation period. In this specific case, the total response time is caused on the one hand by the technical time lag $\tau_{exp}$ and on the other hand by a dynamical and physical delay. This leads us to interpret that the instrumental delay $\tau_{exp}$ between the time the perturbation is generated and the time at which it actually reaches the laser is close to 5~$ns$, which agrees well with our estimations (e.g. length of the cables and optical fibers, free space propagation). After correcting for this technical lag, one can use the SR to analyse the dynamical response of the laser, which has been found to depend on the frequency of the perturbations.

Optically injected lasers have been shown to exhibit multipulse excitability~\cite{multipulse,dolcemascolo2018resonator}, such that the laser emits more than one pulse after a single perturbation. In our system, however, for the range of frequencies and the experimental conditions considered, the success rate was found to be less or equal to one (except when a low spike-detection threshold was used, because in that case, the relaxation oscillations were also counted as response pulses, as discussed in relation to Fig.~\ref{fig:Fig_Th}). Therefore, multipulse excitability was not identified. In our system, for low perturbation frequencies, the oscillations following the ``response pulse'' have a clear shape of relaxation oscillations. As the perturbation frequency increases, the relaxation oscillations gradually disappear (see the time traces shown in Fig.~\ref{fig:Fig_TS}) up to a point in which the laser cannot follow the perturbations and there is a sharp transition from 1:1 locking to higher order locking, where the laser emits less than one pulse per perturbation cycle. It will be interesting for future work to investigate such high order lockings using symbolic analysis, in particular, using ordinal analysis that detects patterns and nonlinear temporal correlations in sequences of events~\cite{bp,mattias,tiana2019}.   

\section{Conclusions}\label{sec:conc}

We have used the ``success rate'' (SR) analysis to study the excitable pulses emitted by an optically injected laser that is periodically perturbed.
The SR method is based on counting the number of response pulses emitted by the laser, within a given time interval, $\Delta T$, after each perturbation.
We have uncovered a lag between the input and output signals, which was traced back to signal transmission in the experimental setup. Once this lag was properly taken into account, the SR provided an accurate identification of the locked and unlocked regimes. We have found frequency regions where the  external perturbation fully controls the excitable laser pulses, such that the laser emits after each perturbation, within the detection window $\Delta T$, a single pulse. 

Taken together, our results show that SR analysis yields relevant information of the dynamics of excitable systems that are periodically perturbed. {{More in general, the success rate is an appropriate measure for studying periodically forced systems, when one needs to identify the optimal conditions under which each small perturbation produces one and only one response. Examples include weak electric periodic stimulation of cardiac tissue for the control of arrhythmia, or of the nervous system for the treatment of brain disorders.}}

\section*{Acknowledgements} \label{sec:acknowledgements}
This work was supported in part by the Spanish Ministerio de Ciencia, Innovacion y Universidades (FIS2015-66503-C3- 2-P). C.M. acknowledges partial support from ICREA ACADEMIA, Generalitat de Catalunya. J.T. acknowledges partial support from the European Commission through the European Fund of Regional Development (FEDER), Convocatoria Excellencia KTT-UPC 2019.

\section*{DATA AVAILABILITY}
The data that supports the findings of this study are available within the article.

\end{document}